\documentclass [12pt] {report}
\usepackage{amssymb}
\newcommand {\eqdef} {\stackrel{\rm def}{=}}
\newcommand {\D}[2] {\displaystyle\frac{\partial{#1}}{\partial{#2}}}

\newcommand {\la} {\lambda}

\newcommand {\de} {\delta}
\newcommand {\prtl} {\partial}
\newcommand {\fr} {\displaystyle\frac}
\newcommand {\wt} {\widetilde}
\newcommand {\be} {\begin{equation}}
\newcommand {\ee} {\end{equation}}
\newcommand {\ba} {\begin{array}}
\newcommand {\ea} {\end{array}}
\newcommand {\bp} {\begin{picture}}
\newcommand {\ep} {\end{picture}}
\newcommand {\bc} {\begin{center}}
\newcommand {\ec} {\end{center}}
\newcommand {\bt} {\begin{tabular}}
\newcommand {\et} {\end{tabular}}
\newcommand {\lf} {\left}
\newcommand {\rg} {\right}

\newcommand {\cF} {{\cal F}}

\newcommand {\cI} {{\cal I}}

\newcommand {\iy} {\infty}
\newcommand {\ses} {\medskip}

\newcommand {\arctg} {\mathop{\rm arctg}\nolimits}

\newcommand {\bR} {{\bf R}}

\newcommand {\bhR} {{\bf\hat R}}

\newcommand{\mhR}{\mbox{$|{\bf\hat R}|$}}
\newcommand{\mwR}{\mbox{$|{\bf\widetilde R}|$}}
\newcommand{\mR}{\mbox{$|{\bf R}|$}}
\newcommand {\cE} {{\cal E}}
\newcommand {\cP} {{\cal P}}
\newcommand {\bibit} {\bibitem}
\newcommand {\nin} {\noindent}
\newcommand {\set}[1] {\mathbb{#1}}

\def\2#1#2#3{{#1}_{#2}\hspace{0pt}^{#3}}
\def\3#1#2#3#4{{#1}_{#2}\hspace{0pt}^{#3}\hspace{0pt}_{#4}}
\newcounter{sctn}
\def\sec#1.#2\par{\setcounter{sctn}{#1}\setcounter{equation}{0}
                  \noindent{\bf\boldmath#1.#2}\bigskip\par}

\begin {document}

\begin {titlepage}

\vspace{0.1in}

\begin{center}
{\Large Finsleroids Reflect Future-Past Asymmetry of Space-Time}\\
\end{center}

\vspace{0.3in}

\begin{center}

\vspace{.15in}
{\large G.S. Asanov\\}
\vspace{.25in}
{\it Division of Theoretical Physics, Moscow State University\\
117234 Moscow, Russia\\
(e-mail: asanov@newmail.ru)}
\vspace{.05in}

\end{center}

\begin{abstract}

The Finslerian extension of the Euclidean metric is proposed and
studied under
rigorous conditions that the associated indicatrix is regular and convex.
The relativistic pseudo-Euclidean metric is extended,
too. The extensions show distinct violation of the $T-$parity,
so that the future-past asymmetry of the
physical world can stem from the $T-$asymmetry of the Finslerian indicatrix.
Indeed the idea should compel much attention of physicists.

\end{abstract}

\end{titlepage}

\vskip 1cm

\sec 1. Introduction
\bigskip

Let us pose the question:

\it Is there a handy possibility to continue the Euclidean metric into the
Finslerian domain such that the Euclidean sphere, $S$, will go over into
a regular and convex closed surface, an indicatrix $\cI$\rm?

On analyzing  the query
\be
S\stackrel{Finsler\,way}{\Longleftrightarrow}\cI\qquad -\qquad ?
\ee
we should note, first of all, that the surface $\cI$
cannot be ``spherical-symmetric", for the choice of a sphere for
the indicatrix
will immediately lead back to the precise-Euclidean metric.

\it Whence the indicatrix $\cI$ should be asymmetric at least in a
single direction\rm. In this respect, the simplest assumption is that
indicatrix is a (hyper)surface of
revolution around  a preferred
direction, to be called conventionally \it the
\rm $T-$\it direction\rm.

In what follows, we would like to propose and study the corresponding
Finslerian metric function obtained under a convenient condition that the
indicatrix is a space of constant positive curvature. We call the respective
indicatrices $\cI$ \it the Finsleroids.\rm

For the spaces under study, the indicatrix equation is essentially non-linear,
so that the generatrix equation
\be
T_{\rm Finslerian}=T(g;\mR),
\ee
where $g$ is the characteristic Finslerian parameter,
is defined only implicitly. This is in contrast to the
ordinary root dependence
\be
T_{\rm Euclidean}=\sqrt{1-\mR^2}
\ee
or
\be
T_{\rm pseudo-Euclidean}=\sqrt{1+\mR^2}.
\ee
It occurs, however, that differentiations of
the indicatrix equations yield sufficiently transparent equations which
provide a convenient basis to study any peculiar feature of any given
Finsleroid.
We shall follow this method.

For fundamentals of Finsler Geometry, the reader is referred to [1-7].
In various respects, the present work continues our previous papers [8-12].

The organization of the paper is as follows.

On introducing  the implied initial definitions in Section 2, we devote
Section 3 to deriving the relations which are sufficient
to prove the regularity and convexity of Finsleroids.
In Section 4, we expose explicitly the relevant diffeomorphic spherical map
\be
S\stackrel{\tau}{\Longleftrightarrow}\cI
\ee
which justifies the above query (1.1) in positive. A direct way of
finding the associated Hamiltonian function in an explicit way is proposed in
Section 5, which enables us in Section 6
to understand that the corresponding Co-Finsleroid (the figuratrix)
is $gT-$conjugate to the Finsleroid (to the indicatrix). After that, in Section 7, we formulate the
relativistic counterpart of our theory, in which case $T$ has the
meaning of a physical time proper.
In the last Section 8 we sum up some straightforward ways of physical
applications of the extensions obtained.
\ses

\sec 2. Initial definitions

Suppose we are given an
($N-1$)-dimensional Euclidean space $E_{N-1}$ with local Euclidean coordinates
$\{R^a\}$ and with the Euclidean metric tensor $e=:\{e_{ab}=\delta_{ab}\}$,
where $\delta$ stands for the Kronecker symbol.
Let us form an $N$-dimensional topological-Euclidean space $\cE$ to be the product
\be
\cE=E_{N-1}\times{\set R},
\ee
where ${\set R}$ is the real line, and
use a canonical coordinate $T$ in ${\set R}$ to
decompose the vectors $R\in\cE$:
\be
R=\{R^p\}=\{R^0=T,R^a\}.
\ee
The indices $(a,b,\dots)$ and
$(p,q,\dots)$ will be specified over the
ranges $(1,\dots,N-1)$ and $(1,\dots,N)$, respectively.
In the vector notation, we have
\be
\bR=\{R^a\}=\{R^1,\dots,R^{N-1}\}
,\qquad \mR=\sqrt{\delta_{ab}R^aR^b},
\ee
and
\be
R=\{T,\bR\}.
\ee
We put
\be
w^a=R^a/T, \quad w=\mR/T,
\quad w_a=w^a.
\ee
The variable $w$
is compatible with the whole definition domain
\be
w\in(-\iy,\iy).
\ee
Throughout this
paper, vector indices are up, co-vector indices are down, and
repeated up--down
indices are automatically summed; $N=4$ in the proper space-time context.

Given a parameter $g$ ranged over
\be
-2<g<2,
\ee
let us introduce the convenient notation
\be
h=\sqrt{1-\fr14g^2},   \qquad r=\fr1h,
\ee
\be
G=g/h,
\ee
together with the characteristic
quadratic form
\be
B(g;R)=\mR^2+g\mR|T+T^2
\ee
whose discriminant is
\be
D=-4h^2<0.
\ee

In terms of this notation, we propose the Finslerian metric function:
\be
K(g;R)=\sqrt{B(g;R)}\,j(g;R),
\ee
where
\be
j(g;R)=\exp\lf(\fr12G(\fr{\pi}{2}-\arctg\fr{2\mR+gT}{2hT})\rg),
\qquad if \quad T\ge 0,
\ee
\medskip
\be
j(g;R)=
\exp\lf(-\fr12G(\fr{\pi}{2}+\arctg\fr{2\mR+gT}{2hT})\rg),
\qquad if \quad T\le 0,
\ee
and
\be
j(g;R)=1, \qquad if \quad T=0.
\ee
Under these conditions, we call the Minkowskian space $\{\cE,K\}$ {\it the
$\cE_{PD}$-space}:
\be
\cE_{PD}=\{\cE=E_{N-1}\times{\set R};K(g;T,\bR);g\}.
\ee

We have
\be
K(g;-T,\bR)\ne K(g;T,\bR),\qquad unless \quad g=0.
\ee
Instead, the function $K$ shows the property of
$gT$-\it parity\rm
\be
K(-g;-T,\bR)=K(g;T,\bR)
\ee
and the property of $\cP$-\it parity\rm
\be
K(g;T,-\bR)=K(g;T,\bR).
\ee

It is frequently convenient to rewrite the representation (2.12) in the form
\be
K(g;R)=|T|V(g;w)
\ee
with the generating function
\be
V(g;w)=\sqrt{Q(g;w)}\,j(g;w),
\ee
where $Q(g;w)$ abbreviates $B(g;R)/T^2$, so that
\be
Q(g;w)=1+gw+w^2.
\ee
We directly obtain
\be
V'=wV/Q, \quad V''=V/Q^2, \qquad
j'=-\fr12gj/Q,
\ee
where the prime ($'$) denotes the differentiation with respect to~$w$.
Using Eqs. (2.19)--(2.23) in the Finslerian
rule
$R_p=\fr12\prtl K^2\Big/\prtl R^p$
yields
\be
R_a=w_aVK/Q,\quad
R_0=(1+gw)VK/Q.
\ee
\bigskip

\sec 3. Shape of Finsleroid
\bigskip

The metric function (2.12) defines an $(N-1)$-dimensional
indicatrix hypersurface according to the equation
\be
K(g;,T,\mR)=1.
\ee
We call this particular hypersurface \it the Finsleroid\rm,
to be denoted as $\cF^{PD}_g$.

From (2.12)-(2.14) it follows directly that
\be\mR_{\big|_{T=0}}=1.
\ee
Also
\be
\lf.T\rg|_{\mR=0}=T_1(g),
\quad when \quad T<0;
\qquad
\lf.T\rg|_{\mR=0}=T_2(g),
\quad when \quad T>0,
\ee
where
\be
T_1(g)=-e^{G\pi/4}\exp\Bigl[\fr G2\arctg{\fr G2}\Bigr] \qquad and
\qquad
T_2(g)=e^{-G\pi/4}\exp\Bigl[\fr G2\arctg{\fr G2}\Bigr].
\ee
 The equation (3.1) cannot be resolved for the function
\be
T=T(g;\mR)
\ee
in an explicit form, because of a complexity of the right-hand part of
Eq. (2.12). Nevertheless, differentiating the identity
\be
K(g;T(\mR),\mR)=1.
\ee
yields the simple result
\be
\fr{dT}{d\mR}=-\fr\mR{T+g\mR}
\ee
which just entails
\be
\fr{d^2T}{d\mR^2}=-\fr{B(g;R)}{(T+g\mR)^3}<0.
\ee
\bigskip
We also get
\be
\lf.\fr{dT}{d\mR}\rg|_{\mR=0}=0,\qquad
\fr{dT}{d\mR}\mathop{\Longrightarrow}\limits_{T\to+0}-\fr1g.
\ee

Inversely, for the function
\be
\mR=\mR(T)
\ee
we obtain
\be
\fr{d\mR}{dT}=-\fr{T+g\mR}{\mR}
\ee
and
\be
\fr{d^2 \mR}{dT^2}=-\fr{B(g;R)}{\mR^3}<0.
\ee
We have
\be
\fr{d\mR}{d T}>0, \quad if \quad  T<-g\mR; \qquad and \quad
\fr{d\mR}{dT}<0, \quad if \quad  T>-g\mR.
\ee
Also,
\be
\fr{d\mR}{d T}=0, \quad if \quad  T= T^* \quad with \quad  T^*=-g\mR.
\ee
Inserting this $T^*$ in (3.1) yields
\be
T^*=f(g)
\ee
with
\be
f(g)=-g\exp\lf(\fr{G}2(\fr{\pi}{2}-\arctg\fr{2-g^2}{2gh})\rg),
\ee
and for the function
\be
k(g)\eqdef\mR(T^*)
\ee
we obtain merely
\be
k(g)=\exp\lf(\fr{G}2(\fr{\pi}{2}-\arctg\fr{2-g^2}{2gh})\rg).
\ee

The above formulae, particularly the negative sign of
the second derivative (3.8), are useful to apply when verifying the following

THEOREM 1. \it The Finsleroid $\cF_g^{PD}$ is closed, regular, locally-convex
and convex.\rm
\bigskip

\sec 4. Spherical map of Finsleroid
\bigskip

Let us perform in the space $\cE_{PD}$ the nonlinear transformation given by
the functions
\be
\wt R^q=\tau^q(g;R)
\ee
with
\be
\tau^0=(R^0+\fr12g\mR)j(g;R^0,\mR)r(g)
, \qquad \tau^a=R^aj(g;R^0,\mR),
\ee
and call the result \it the \rm$\tau$-\it transformation\rm.
Inserting these functions in an Euclidean metric function
\be
S(\wt R)\eqdef\sqrt{(\wt R^0)^2+\mwR^2}
\ee
yields the remarkable identity
\be
K(g;R)=S(\wt R)/r(g),
\ee
where $K(g;R)$ is exactly the Finslerian metric function (2.12); $h(g)$ and
$r(g)$ are the functions that were defined in Eq. (2.8).
Therefore we have

THEOREM 2. \it The $\tau$-transformation turns over the Finsleroid $\cF_g^{PD}$
into the sphere $S_{r(g)}$ of radius $r(g)$.\rm

An attentive consideration shows that the functions written out in Eq. (4.2)
are smooth of at least class $C^2$ over all the definition range
\be
(R^0)^2+\mR^2>0.
\ee
Whence \it the $\tau$-transformation of the Finsleroid $\cF_g^{PD}$
into the sphere $S_{r(g)}$ is a diffeomorphism. \rm

By the help of (4.1) and (4.2) we find
\be
w=\mwR/\wt I,
\ee
where
\be
\wt I=\wt I(g;\wt R)=h(g)\wt R^0-\fr12g\mwR.
\ee
Additional direct calculations lead to the relation
\be
Q(g;w)=\fr{(\wt R^0)^2+\mwR^2}{r^2(g)\wt I^2(g;\wt R)}
\ee
which, when used together with the redefinition
\be
\wt j(g;\wt R)\eqdef j(g;R(g;\wt R))=j\Bigl(g;\fr{\mwR}{\wt I}\Bigr),
\ee
enables us to inverse the transformations (4.1)--(4.3):
\be
R^p=\la^p(g;\wt R),
\ee
where
\be
\la^0=\wt I(g;\wt R)/\wt j(g;\wt R), \quad \la^a=\wt R^a/\wt j(g;\wt R).
\ee

The functions (4.2) are homogeneous of degree 1 with respect to
~$R$:
\be
\tau^q(g;bR)=b\tau^q(g;R), \qquad b>0,
\ee
from which it follows that the identity
\be
\tau_p^q(g;R)R^p=\wt R^q
\ee
holds for the derivatives
\be
\tau_p^q(g;R)\eqdef\D{\tau^q(g;R)}{R^p}.
\ee
The simple representations
\be
\tau_0^0=\lf(1-\fr12gwEQ^{-1}\rg)j/h,
\ee\be
\tau_a^0=\fr12g(w_a/w)(E-Q)Q^{-1}j/h,
\ee\be
\tau_0^a=-\fr12gww^aQ^{-1}j,
\quad \tau_b^a=j\de_b^a+\fr12g(w^aw_b/w)Q^{-1}j
\ee
are obtained, where
\be
E=1+\fr12gw.
\ee
The determinant is equal to
\be
\det(\tau_p^q)=rj^N>0.
\ee
The relations
$$
\tau_b^aw^b=jw^a(E-w^2)Q^{-1}, \quad
\tau_c^a\tau_d^br^{cd}=j^2\lf[\delta^{ab}+g(w^aw^b/wQ)+\fr14g^2(w^aw^b/Q^2)\rg]
$$
are convenient to take into account in process of calculations involving
coefficients $\{\tau_p^q\}$.

Similarly to (4.12), we get
\be
\la^p(g;b\wt R)=b\la^p(g;\wt R), \qquad b>0,
\ee
which entails the identity
\be
\la_q^p(g;\wt R)\wt R^q=R^p
\ee
for the derivatives
\be
\la_q^p(g;\wt R)\eqdef\D{R^p(g;\wt R)}{\wt R^q}.
\ee

Let us find the transform
\be
n^{pq}(g;\wt R)\eqdef\tau_r^p(g;R)\tau_s^q(g;R)
g^{rs}(g;R).
\ee
of the Finslerian metric tensor $g^{rs}(g;R)$ (associated to the Finslerian
metric function (2.12)) under the $\tau$-transformation.
By the help of Eqs. (4.12)-(4.18),
we get after rather lengthy calculations the following simple result:
\be
n^{pq}=\delta^{pq}+\fr14G^2l^pl^q, \quad
n_{pq}=\delta_{pq}-\fr14g^2l_pl_q
\ee
($n_{pr}n^{qr}=\de_p^q$; $G$ is given by Eq. (2.9)), and
\be
\det(n_{pq})=h^2,
\ee
where
\be
l^p=\wt R^p/S(\wt R)=l_p
\ee
are Euclidean unit vectors, obeying the rules
$$
l^pl_q=1, \quad
n_{pq}l^q=h^2l_p, \quad n^{pq}l_q=l^p/h^2, \quad
n_{pq}l^pl^q=h^2.
$$
Inversing (4.23) reads
\be
g_{pq}(g;R)=n_{rs}(g;\wt R)\tau_p^r(g;R)\tau_q^s(g;R).
\ee
We call the tensor $\{n\}$ with components  (4.24) \it
the quasi-Euclidean metric tensor\rm.

Thus we have proven

THEOREM 3. \it The $\tau$-transformations turns over the
Finslerian metric tensor
of the space
$\cE_{PD}$ under study in the quasi-Euclidean metric tensor
in accordance with \rm(4.23) \it or \rm(4.27).

The $\tau$-transformation defines obviously \it the \rm $\tau$-\it map\rm
\be
\cF_g^{PD}\stackrel{\tau}{\Longleftrightarrow}S_{r(g)}
\ee
which is a diffeomorphism.
Vice versa, because the $\tau$-transformations are homogeneous,
according to Eqs. (4.12) and (4.20), the knowledge of the $\tau$-map
can be applied to restore totally the $\tau$-transformation.
\bigskip

\sec 5. Associated Hamiltonian function
\bigskip

To go over from the space (2.16) to its dual counterpart, $\hat\cE$,
we ought to introduce the co-versions of Eqs. (2.2)-(2.6): $\hat R\in\hat\cE$
and
\be
\hat R=\{R_p\}=\{R_0=\hat T,R_a\},
\ee
\bigskip
\be
\bhR=\{R_1,\dots,R_{N-1}\}, \qquad
\mhR=\sqrt{\delta^{ab}R_aR_b},
\ee
\bigskip
\be
\hat R=\{\hat T, \bhR\},
\ee
\bigskip
\be
p_a=R_a/\hat T, \quad p=\mhR/\hat T,  \quad p^a=p_a,
\ee
where $p\in(-\iy,\iy),$ and consider the quadratic form
\be
\hat B(g;\hat R)=
\mhR^2-g\hat T\mhR +(\hat T)^2
\ee
conjugated to the basic form (2.10).

To find the Hamiltonian function $H$ associated to the Finslerian metric
function (2.12), we should resolve the equation set (2.24) with respect to
the variables $\{R^p\}$ to construct
\be
H(g;\hat R)\eqdef K(g;R)
\ee
(see the respective general homogeneous Hamilton-Jacobi theory in [13-14]).
This procedure yields
\be
H(g;\hat R)=
\sqrt{\hat B(g;\hat R)}\,\hat j(g;p)
\ee
where
\be
\hat j(g;p)=\exp\lf(\fr12G(-\fr{\pi}2+\arctg\fr{2\mhR-g\hat T}{2h\hat T}\rg),
\qquad if \quad \hat T\ge 0,
\ee
\medskip
\be
\hat j(g;p)=\exp\lf(\fr12G(\fr{\pi}2+\arctg\fr{2\mhR-g\hat T}{2h\hat T}\rg),
\qquad if \quad \hat T\le 0.
\ee
and
\be
\hat j(g;p)=1
, \qquad if \quad \hat T=0.
\ee

We observe that
\be
H(g;-\hat T,\bhR)\ne H(g;\hat T,\bhR),\qquad unless \quad g=0.
\ee
At the same time, quite similarly to the properties (2.18) and (2.19),
the function $H$ shows the property of $g\hat T$-\it parity\rm
\be
H(-g;-\hat T,\bhR)=H(g;\hat T,\bhR)
\ee
and the property of $\hat \cP$-\it parity\rm
\be
H(g;\hat T,-\bhR)=H(g;\hat T,\bhR).
\ee

In an alternative way, we write
\be
H(g;\hat R)=|\hat T|W(g;p)
\ee
with
\be
W(g;p)=\sqrt{\hat Q(g;p)}\,\hat j(g;p),
\ee
where
\be
\hat Q(g;p)=1-gp+p^2\equiv \hat B(g;\hat R)/(\hat T)^2.
\ee
Similarly to (2.23), we obtain
\be
W'=pW/\hat Q, \quad W''=W/(\hat Q)^2,\qquad
\hat j'=\fr12g\hat j/\hat Q,
\ee
which entails for the components of the contravariant vector
$R^p=\fr12\prtl H^2\Big/\prtl R_p$ the following result:
\be
R^a=p^aWH/\hat Q,\quad
R^0=(1-gp)WH/\hat Q.
\ee

The identities
\be
\hat j(g;p)=1/j(g;w),
\ee
\bigskip
\be
Q(g;w)=\fr{\hat Q(g;p)}{(1-gp)^2},
\ee
and
\be
V^2W^2=Q\hat Q
\ee
hold fine.

To verify that the representations (4.18) solve the set of equations (2.24),
it is easy to note that Eqs.~(5.4) and (2.24) entail the equality
\be
p_a=w_a/(1+gw),
\ee
which inverse is
\be
w_a=p_a/(1-gp).
\ee
In this way we find
\be
p=\fr w{1+gw}, \quad w=\fr p{1-gp}, \quad 1+gw=\fr1{1-gp}.
\ee
When the last relations are used in the definition (2.21) for the function $Q$,
the identities (5.19) and (5.20) are obtained, whereupon we take into
account the second part of Eq. (2.24) to obtain the equality
\be
\hat T=(1+gw)K^2/QT,
\ee
which entails
\be
T=\fr{1+gw}{Q(g;w)}\,\fr{K^2}{\hat T}=\fr{1-gp}{\hat Q(g;p)}\,\fr{K^2}{\hat T}.
\ee
The conclusive step is to insert (5.26) in the right-hand part of (2.12).
\bigskip

\sec 6. Shape of Co-Finsleroid
\bigskip

The Hamiltonian function (5.7) gives rise an ($N-1)$-dimensional figuratrix to be
the hypersurface defined by the equation
\be
H(g;\hat T,\mhR)=1.
\ee
We call this particular hypersurface \it the Co-Finsleroid\rm, to be denoted as
$\hat\cF^{PD}_g$.

Evaluating the functions (5.7)-(5.9) at $\hat T=0$ yields
\be
\mhR_{\big|_{\hat T=0}}=1.
\ee
Also
\be
\lf.\hat T\rg|_{\mhR=0}=-T_1(g),
\quad when \quad \hat T>0; \qquad
\lf.\hat T\rg|_{\mhR=0}=-T_2(g),
\quad when \quad \hat T<0,
\ee
where $T_1(g)$ and $T_2(g)$ are exactly the functions given by Eq. (3.4).

Differentiating the identity
\be
H(g;\hat T(\mhR),\mhR)=1
\ee
lieds to the simple result
\be
\fr{d\hat T}{d\mhR}=-\fr \mhR{\hat T-g\mhR},
\ee
from which it follows that
\be
\fr{d^2\hat T}{d\mhR^2}=-\fr{\hat B(g;R)}{(\hat T-g\mhR)^3}<0.
\ee
\bigskip
We also get
\be
\lf.\fr{d\hat T}{d\mhR}\rg|_{\mhR=0}=0,\qquad
\fr{dT}{d\mhR}\mathop{\Longrightarrow}\limits_{\hat T\to+0}\fr1g.
\ee

Inversely,
we obtain
\be
\fr{d\mhR}{d\hat T}=-\fr{\hat T-g\mhR}{\mhR}
\ee
and
\be
\fr{d^2 \mhR}{d\hat T^2}=-\fr{\hat B(g;R)}{\mhR^3}<0.
\ee
\bigskip
We have
\be
\fr{d\mhR}{d\hat T}>0, \quad if \quad \hat T<g\mhR; \qquad and \quad
\fr{d\mhR}{d\hat T}<0, \quad if \quad \hat T>g\mhR.
\ee
\bigskip
Also,
\be
\fr{d\mhR}{d\hat T}=0, \quad if \quad \hat T=\hat T^* \quad with \quad \hat T^*=g\mhR.
\ee
Inserting this $\hat T^*$ in (6.1) yields
\be
\hat T^*=\hat f(g)
\ee
with
\be
\hat f(g)=g\exp\lf(\fr{G}2(\fr{\pi}{2}+\arctg\fr{2-g^2}{2gh})\rg)\equiv f(-g)
\ee
(cf. Eq. (3.17)).
\ses

The above calculations enable us to obtain
\bigskip

THEOREM 4. \it The Co-Finsleroid $\hat\cF^{PD}_g$ is closed, regular,
locally-convex and convex.
The  $\cF^{PD}_g$-hyperboloid and
the $\hat\cF^{SR}_g$-co-hyperboloid mirror one another
under the $g$-reflection: \rm
\be
\cF^{PD}_g\stackrel{g\longleftrightarrow -g}
{\Longleftrightarrow}\hat\cF^{PD}_{-g}.
\ee
\bigskip

\sec 7. Shape of $\cF^{SR}_g$-hyperboloid
\bigskip

The special-relativistic Finslerian metric function
\be
F_{SR}(g;R)=\lf|T+g_-|{\bf R}|\rg|^{G_+/2}\lf|T+g_+|{\bf R}|\rg|^{-G_-/2}
\ee
can be adduced by the Hamiltonian function
\be
H_{SR}(g;\hat R)=\lf|\hat T-\fr{|{\bhR}|}{g^+}\rg|^{G^+/2}
\lf|\hat T-\fr{|{\bhR}|}{g^-}\rg|^{-G^-/2}
\ee
(see [10-12]). In this case we ought to replace the definition (2.8) by
\be
\qquad h\eqdef\sqrt{1+\fr14g^2}
\ee
and use the notation
\be
g_+=-\fr12g+h, \qquad g_-=-\fr12g-h,
\ee
\medskip
\be
g^+=1/g_+=-g_-,  \qquad  g^-=1/g_-=-g_+,
\ee
\medskip
\be
g^+=\fr12g+h, \qquad g^-=\fr12g-h.
\ee
\medskip
\be
G_+=g_+/h, \quad G_-=g_-/h,
\ee
\medskip
\be
G^+=g^+/h, \quad G^-=g^-/h.
\ee

The associated indicatrix  equation
\be
F_{SR}(g;T,\mR)=1
\ee
defines what we call \it the $\cF^{SR}_g$-hyperboloid\rm.
From (7.1) we get
\be
\mR_{\big|_{T=0}}=c(g),
\ee
where
\be
c(g)=(-g_-)^{-G_+/2}(g_+)^{G_-/2},
\ee
and also
\be
\lf.T\rg|_{\mR=0}=1,
\quad when \quad T>0; \qquad
\lf.T\rg|_{\mR=0}=-1,
\quad when \quad T<0.
\ee
Differentiating Eq. (7.9)
yields the simple result
\be
\fr{dT}{d\mR}=\fr\mR{T-g\mR}
\ee
which entails
\be
\fr{d^2T}{d\mR^2}=\fr{B(g;R)}{(T-g\mR)^3}>0.
\ee
\bigskip
We observe that
\be
\lf.\fr{dT}{d\mR}\rg|_{\mR=0}=0,\qquad
\fr{dT}{d\mR}\mathop{\Longrightarrow}\limits_{T\to+0}-\fr1g.
\ee

Inversely,
\be
\fr{d\mR}{dT}=\fr{T-g\mR}{\mR}
\ee
leads to
\be
\fr{d^2 \mR}{dT^2}=\fr{B(g;R)}{\mR^3}>0.
\ee
We have
\be
\fr{d\mR}{dT}>0, \quad if \quad T>g\mR; \qquad and \quad
\fr{d\mR}{dT}<0, \quad if \quad T<g\mR.
\ee
Also,
\be
\fr{d\mR}{dT}=0, \quad if \quad T=T^* \quad with \quad T^*=g\mR.
\ee
Inserting this $T^*$ in (7.1) yields
\be
T^*=s(g)
\ee
with
\be
s(g)=gz(g),
\ee
where
\be
z(g)=(g_+)^{-G_+/2}(-g_-)^{G_-/2},
\ee
and also
\be
R^*\eqdef \mR(T^*)=z(g).
\ee

Quite similar formulae are obtainable for the
\it $\hat\cF^{SR}_g$-co-hyperboloid \rm
defined by the figuratrix equation
\be
H_{SR}(g;\hat R)=1.
\ee

From Eqs. (7.1)-(7.22) we derive
the following counterpart of the $\{PD\}$-Theorems
1 and 2:

THEOREM 5. \it The  $\cF^{SR}_g$-hyperboloid is everywhere regular and
locally-convex.
The same conclusion is applicable to the
$\hat\cF^{SR}_g$-co-hyperboloid. These two hypersurfaces,
the  $\cF^{SR}_g$-hyperboloid and
the $\hat\cF^{SR}_g$-co-hyperboloid,
mirror one another under the $g$-reflection: \rm
\be
\cF^{SR}_g\stackrel{g\longleftrightarrow -g}
{\Longleftrightarrow}\hat\cF^{SR}_{-g}.
\ee
\bigskip

\sec 8. Conclusion: New Ways for Finslerian Physics?

\ses

There exists rather huge literature (see [15]) about
possible violation of the Special Theory of Relativity (STR).
In several instances, the authors proposed sensitive ways to test
experimentally how well Lorentz invariance is obeyed in Nature (see [16-18]).
Nevertheless, one cannot say that the attempts made
were conclusive. In fact, much more informatiion that were obtained
is needed actually to put reliable limits on Lorentz noninvariance.

To consider departures from the STR, the standard practice was
to modify the Lorentz transformations while leaving the
pseudoEuclidean metric intact. However, we know (and teach students!)
that the Lorentz transformations stem directly from the choice of
the latter metric because they are playing actually the role of invariance
transformations.
Therefore, to investigate possible violation of Lorentz transformations in
self-consistent way, one should modify the Lorentz transformations
in conjunction with a due Finslerian extension of the metric.
Accordingly, to rectify the practice, we propose
to follow the concise Finslerian approach
outlined in Sec.7 (and in the previous papers [8-12]), in which
the characteristic parameter $g$ measures the degree of violation
of the Finslerian metric function \it simultaneously with \rm the degree of violation of
the Lorentz transformations.

Our approach is everywhere compatible with the ordinary believe that
``the laws of physics are invariant under spatial rotations".
At the same time, the resultant Finslerian framework manifests
the $T-$violation, so that we ought to conclude that \it the parameter
$g$ measures also the degree of violation of the T-parity, and
hence the CP-parity. \rm Whence searches for the $CP-$violation
(which are many; see, e.g., [19-20]) can, in principle,
put interesting limits on the magnitude of the parameter $g$.

The theory of relativistic quantum fields adheres often
to the so-called ``Euclidean turn"
which go over the theory in the Euclidean sector (see, e.g. [19-24]; such
a sector is often used to consider the confinement of quarks).
In the context of this, the positive-definite Finslerian metric function
described in Secs. 2-6 may serve to give the base to continue ``the Euclidean
theory of quantum fields" in the Finslerian domain. Again,
 the continuation proves to be ``$T-$asymmetric", for the Finsleroids are not
symmetric under the $T-$ reflection, - instead they are mirror-symmetric
under the $gT$-type reflections (cf. Eq. (2.18)).

In general, the Finslerian approach outlined above seems to follow
the proud thesis: \it``Sometimes a physical motivation precedes a mathematical
theory, but it is not always so" \rm  [R.S.Ingarden, [25], p.213].
\bigskip

\def\bibit[#1]#2\par{\rm\noindent\parskip1pt
                     \parbox[t]{.05\textwidth}{\mbox{}\hfill[#1]}\hfill
                     \parbox[t]{.925\textwidth}{\baselineskip11pt#2}\par}

\nin{\bf References}
\bigskip

\bibit[1] R. S. Ingarden and L. Tamassy: \it Rep. Math. Phys.
\bf32 \rm(1993), 11.

\bibit[2] E. Cartan: \it Les espaces de Finsler, Actualites \rm 79, Hermann,\quad
Paris 1934.

\bibit[3] H. Rund: \it The Differential Geometry of Finsler spaces, \rm
Springer-Verlag, Berlin 1959.

\bibit[4] R. S. Ingarden: \it Tensor \bf30 \rm(1976), 201.

\bibit[5] G. S. Asanov: \it Finsler Geometry, Relativity and Gauge Theories, \rm
D.~Reidel Publ. Comp., Dordrecht 1985.

\bibit[6] D.~Bao, S. S. Chern, and Z. Shen (eds.): \it Finsler Geometry \rm
(Contemporary Mathematics, v.~196), American Math. Soc., Providence 1996.

\bibit[7] D.~Bao, S. S. Chern, and Z. Shen: \it An
Introduction to Riemann-Finsler Geometry
,\quad\rm Springer, N.\.Y, Berlin, 2000.

\bibit[8] G. S. Asanov: \it Aeq. Math.\bf49 \rm(1995), 234.

\bibit[9] G. S. Asanov: \it Rep. Math. Phys. \bf39 \rm(1997), 69;
\bf 41 \rm (1998), 117; \bf 42 \rm(1998), 273.

\bibit[10] G. S. Asanov: \it Rep. Math. Phys. \bf 45 \rm(2000), 155.

\bibit[11] G.S. Asanov: \it Moscow University Physics Bulletin
 \bf49\rm(1) (1994),~18; \bf51\rm(1) (1996),~15; \bf51\rm(2)
 (1996)~6; \bf51\rm(3) (1996)~1; \bf53\rm(1) (1998),~15.

\bibit[12] G.S. Asanov: \it Moscow University Physics Bulletin
\bf49\rm(2) (1994),~11.

\bibit[13] H. Rund: \it The Hamiltonian--Jacobi Theory in the Calculus of
Variations,\quad\rm Krieger, N.\,Y. 1973.

\bibit[14] H. Rund: Questiones Math., \bf1 \rm(1976), 29.

\bibit[15] Proceedings of Conference ``Physical Interpretation of
Relativity Theory", September 15-20, London, Sunderland, 2000.

\bibit[16] S.Coleman and S.L.Glashow: Cosmic ray and neutrino tests of
special relativity
\it Harvard University Report \rm No. HUTP-97/A008, hep-ph/9703240, 1-5.
; High-energy tests of Lorentz invariance,
\it Phys. Rev. \bf D59 \rm N 7 (1999),  11608.

\bibit[17] S.L.Glashow, A.Halprin, P.L.Krastev, C.N.Leung, and Pantaleone:
Remarks on  neutrino tests of special relativity. Phys. Rev.D \bf56
\rm ,N 4 (1997), 2433-2434.

\bibit[18] G.S.Asanov: Can neutrinos and high-energy particles test
Finsler metric of space-time?
arXiv:hep-ph/0009305, 2000.

\bibit[19] C.Itzykson and J.-B.Zuber: \it Quantum Field Theory
,\quad\rm McGraw-Hill, 1980.

\bibit[20] S.Weinberg: \it The Quantum Theory of Fields,\quad\rm Syndicate
of the University of Cambridge, v.1, 1995; v.2. 1998.

\bibit[21] J.Schwinger: \it Proc. Nat. Acad. Sci.\bf44 \rm(1958), 956.

\bibit[22] K.Osterwalder and R.Schrader: \it Phys. Rev. Lett.\bf29
\rm(1972), 1423.

\bibit[23] K.Osterwalder and R.Schrader: \it Commun. Math. Phys. \bf31
\rm(1973), 83; \bf42 \rm(1975), 281.

\bibit[24] E.R.Speer: \it J.Math.Phys.\bf9 \rm(1968), 1404.

\bibit[25] R. S. Ingarden: In: \it Finsler Geometry \rm (Contemporary
Mathematics, v.~196), American Math. Soc., Providence 1996, pp.~213--223.

\end {document}